\documentclass[twocolumn,10pt]{article}
\input epsf
\input{epsf.sty}
\begin{document}

\title{\Large Entropy, Vortex Interactions and the Phase Diagram of Heavy-Ion 
Irradiated Bi$_{2}$Sr$_{2}$CaCu$_{2}$O$_{8+\delta}$}
 
\twocolumn[
\hsize\textwidth\columnwidth\hsize
\csname@twocolumnfalse\endcsname


\author {C.J. van der Beek$^{1}$, M. Konczykowski$^{1}$, R.J. 
Drost$^{2}$, P.H. Kes$^{2}$, \\ A.V. Samoilov$^{3}$, N.Chikumoto$^{4}$, S. 
Bouffard$^{5}$, M.V. Feigel'man$^{6}$ \\
\footnotesize \em $^{1}$Laboratoire des Solides Irradi\'{e}s, Ecole 
Polytechnique, 91128  Palaiseau, France \\
\footnotesize \em $^{2}$Kamerlingh Onnes Laboratorium, P.O. Box 9506,  2300  RA Leiden, 
 the Netherlands \\
\footnotesize \em $^{3}$Condensed Matter Physics 114-36, California Institute of Technology, 
 Pasadena  CA 91125, U.S.A. \\
\footnotesize \em $^{4}$Superconductivity Research Laboratory, ISTEC, Minato-ku, Tokyo 105, 
Japan \\
\footnotesize \em $^{5}$Centre Interdisciplinaire de Recherche avec les Ions Lourds 
 (C.I.R.I.L.),  B.P. 5133, 14040 Caen Cedex, France \\
\footnotesize \em $^{6}$Landau Institute of Theoretical Physics, Moscow, Russia}
 \date{}
	\maketitle 
	\hrulefill

\begin{abstract}
\small Using dynamic and thermodynamic magnetization measurements, we analyze the phase diagram of 
Bi$_{2}$Sr$_{2}$CaCu$_{2}$O$_{8+\delta}$ single crystals containing amorphous columnar defects 
created by heavy-ion irradiation. Reversible magnetization experiments yield the respective 
magnitudes of the pinning energy and entropy contributions to the free energy of the vortex 
lattice. It appears that the entropy contribution in the London regime is relatively minor in both 
unirradiated and irradiated crystals, except in the case of high density of columns and 
inductions $B$ that are smaller than the interaction field $H_{int} 
\approx B_{\phi}/6$. The dependence of the entropy contribution on vortex- and 
defect density correlates well with measurements of the irreversibility line
$H_{irr}(T)$, which shows a sharp increase at $H_{int}$.
\end{abstract}
\hrulefill
\vspace{15mm}

]
\small
\section*{\normalsize \bf 1. Introduction}
\pagestyle{myheadings}
\markboth{ \footnotesize \em C.J. vander Beek et al. / Phase diagram of heavy-ion 
irradiated BSCCO}{ \footnotesize \em C.J. vander Beek et al. / Phase diagram of heavy-ion 
irradiated BSCCO}

The enormous enhancement of the critical current density [1-4] and of 
the irreversibility line (IRL) separating the vortex glass and liquid phases [3,4] 
obtained after the introduction of amorphous columnar defects into the layered high-Tc 
superconductor Bi$_{2}$Sr$_{2}$CaCu$_{2}$O$_{8+\delta}$ has spurred numerous investigations. Whereas initial studies 
conducted at temperatures below 20 K seemed to indicate that flux pinning and dynamics in 
this sytem could be described within a simple model describing the interaction between 
single pancake vortices (i.e. the intersections of the vortex lines with the superconducting 
CuO$_{2}$-double layers) and the insulating defects [1,2], later work showed that collective 
effects should be taken into account in order to understand the vortex physics at higher 
temperatures (above 40 K). The field dependence of the sustainable current density [5] 
can only be explained if the in-plane repulsion between pancakes is taken into account. 
Further experiments on the dependence of the screening current j on the angle between 
the applied magnetic field and the linear tracks demonstrated that above 40 K pancakes 
belonging to the same vortex line (''stack'') move in a correlated way [6,7]; the 
stretched-exponential time dependence of the screening current, $j \sim (\ln 
t)^{-1/\mu}$ [4,8], 
supports this finding. Measurements of the c-axis resistivity [9-11] and c-axis critical 
current, using Josephson Plasma Resonance (JPR) [12-14], show a drastic enhancement of 
these quantities when the field is increased above $ B_{\phi}/6$; this has been interpreted in 
terms of pancake vortex alignment and an ensuing ''recoupling transition'' in the 
vortex liquid phase. It was proposed that the recoupling is driven by the competition 
between the attractive interaction between pancakes in different layers and the entropy 
gain obtained by spreading pancakes belonging to the same stack over different columns 
[14-16]. It must be stressed however that the fields under study are (much) larger than 
the crossover field $B_{cr} \equiv \Phi_{0}/\gamma^{2}s^{2}$ (with 
$\gamma \sim 300$ the anisotropy parameter, $s = 15$ nm the 
spacing between CuO2 double layers in the 
Bi$_{2}$Sr$_{2}$CaCu$_{2}$O$_{8+\delta}$ material, and $\Phi_{0} = h/2e$ the 
flux quantum). This means that the interlayer pancake interaction is much weaker than 
the intralayer interaction. A number of authors have stressed the importance of the 
intraplane pancake repulsion in determining the vortex configuration and vortex dynamics 
in the presence of columnar defects [17,18]. In this paper, we analyze the reversible 
magnetization of heavy-ion irradiated Bi$_{2}$Sr$_{2}$CaCu$_{2}$O$_{8+\delta}$ single crystals. We determine 
the relative magnitudes of the free energy gain associated with vortex pinning on the 
irradiation-induced amorphous columnar tracks and the entropy gain associated with 
the possibility of pancake vortices to occupy different tracks. We show how information 
on vortex alignment can be obtained from the reversible magnetization, and how the 
vortex arrangement affects  their dynamics and the phase diagram.

\section*{\normalsize \bf 2. Experimental details}

Different batches of Bi$_{2}$Sr$_{2}$CaCu$_{2}$O$_{8+\delta}$ single crystals were grown using the travelling-solvent 
floating-zone technique at the University of Amsterdam and the University of Tokyo. A number 
of crystals were annealed in air at 800¡C in order to obtain the optimum $T_{c} = 90$ K. Other 
crystals were left as-grown (slightly overdoped) with a $T_{c}$ of 83 K. After imaging of the flux 
penetration by magneto-optics and the removal of defective regions, a number of large pieces 
of size $2 \times 1$ mm$^{2}$ $ \times 20$ mm were cut for the reversible magnetization experiments. Smaller 
pieces of size $800 \times 800 \times 20$ $\mu$m$^{3}$ were prepared for
	AC transmittivity measurements of the irreversibility line (IRL) [8]. The crystals were 
	subsequently irradiated with 5.8 GeV Pb$^{56+}$ ions at the Grand 
	Acc\'{e}l\'{e}rateur National 
	d'Ions Lourds (GANIL) in Caen. The ion beam was directed along the 
	$c$-axis; each ion 
	impact created an amorphous columnar track of radius $c_{0} \approx 3.5$ nm, traversing the
	entire crystal thickness. The ion dose for different crystals varied between 
	$1 \times 10^{9}$ and $2 \times 10^{11}$  cm$^{-2}$, corresponding to dose-equivalent matching fields 
	$B_{\phi} \equiv \Phi_{0}n_{d}$ between 20 mT and 2 T ($n_{d}$ is the columnar defect density). The 
	reversible magnetization was measured in Leiden using a commercial SQUID
	magnetometer. The IRL's presented below

\begin{figure}
		\centerline{\epsfxsize 7.5cm \epsfbox{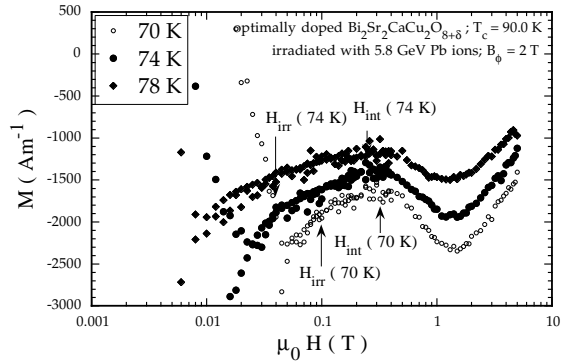}}
		\caption{\small Magnetization of an optimally doped Bi$_{2}$Sr$_{2}$CaCu$_{2}$O$_{8+\delta}$ single crystal 
		grown at the University of Amsterdam, and irradiated with 5.8 GeV Pb ions 
		so as to produce a matching field $B_{\phi} = 2$ T. The arrows denote the irreversibility 
		field $H_{irr}$ as determined from the high temperature onset of a third harmonic 
		response in field-cooled AC transmittivity experiments, as well as the interaction 
		field $H_{int}$ above which intervortex repulsion starts to limit the free energy gain 
		that can be obtained from vortex localization on the columnar defects.}
	\end{figure}
	
    were determined by ac Local Hall 
	Probe Magnetometry [8,19] using an ac field of amplitude 6 Oe and frequency 
	7.753 Hz, applied colinearly with dc field and the crystalline $c$-axis. The 
	irreversibility temperature $T_{irr}(B)$ (or irreversibility field $B_{irr}(T)$) is
	defined as that at which one first observes a third harmonic in the ac 
	screening current when cooling.

\section*{\normalsize \bf 3. Results and Discussion}
\subsection*{\normalsize \it 3.1 Reversible magnetization}

 Figure 1 shows the magnetization of an optimally doped Bi$_{2}$Sr$_{2}$CaCu$_{2}$O$_{8+\delta}$ single crystal 
 with $B_{\phi} = 2$ T. The reversible magnetization Mrev exhibits the usual features found 
 after heavy-ion irradiation [20,21]. Note that the reversible magnetization corresponds 
 to the vortex chemical potential $\mu$, i.e. $M_{rev} = 
 -\partial G/\partial B = -\Phi_{0}-1\partial G/\partial n_{v} = 
 -\Phi_{0}^{-1}\mu$ and is 
 therefore equal to the free energy $\Delta G$ needed to insert (or to extract) a vortex into 
 (from) the vortex lattice. At inductions much less than $B_{\phi}$ all vortices can lower their
 free energy by becoming localized on a defect site, hence the absolute value of the 
 magnetization $|M_{rev}|$ is reduced with respect to that of the usual magnetization in the 
 London model, $M_{rev}^{0} = -\varepsilon_{0}/2\Phi_{0} \ln(\eta 
 B_{c2}/eB)$, by a corresponding amount $(\varepsilon_{0}(T)= 
 \Phi_{0}2/4\pi \mu_{0}\lambda^{2}(T) $
 is the vortex line energy, $\lambda(T) $ is the penetration depth, 
 and $\eta \approx 1$). The fact that this 
 amount is large, {\em i.e.} of the same order of magnitude as the total magnetization, means 
 that vortices are strongly bound to the columnar defects. As the matching field is approached, 
 favourable sites become rare and more and more "interstitial" vortices 
 ({\em i.e.} not localized 
 on a columnar defect) with higher free energy appear. Hence $|M_{rev}|$ increases (at the interaction 
 field, labelled $H_{int}$ in Fig.1). At fields far above $B_{\phi}$, no defect sites are available anymore, 
 all vortices that enter the sample occupy sites in the intercolumn space; hence, the magnetization
 is comparable to $M_{rev}^{0}$. The rounding of the magnetization near $B_{\phi}$ is the result of intervortex
 repulsion, which prohibits the occupation of close-lying columnar defect sites even at fields 
 well below $B_{\phi}$ [18,20], and forces vortices to occupy interstitial sites.
 
The free energy gain manifest at low fields was initially interpreted as being the sole
result of pancake vortex pinning on the columnar tracks. Then, 
$M_{rev}(B <<B_{\phi}) = M_{rev}^{0} + U_{0}/\Phi_{0}$, 
and the pinning energy per unit length $U_{0}$ can be directly obtained 
as the difference between $M_{rev}$
(at low field) and the extrapolation to low fields of the magnetization 
at fields much above $B_{\phi}$. 
The application of this procedure to irradiated crystals with relatively 
low $B_{\phi}$ has been used 
to establish that the main pinning mechanism in heavy-ion irradiated Bi$_{2}$Sr$_{2}$CaCu$_{2}$O$_{8+\delta}$ arises 
from the interaction of the vortex core with the defects [22]. Applying the same method to 
the magnetization of crystals with higher $B_{\phi}$ (as in Fig.1) yields pinning energies that 
exceed any reasonable theoretical prediction. Conversely, in numerical simulations of 
column occupation [18] one has to assume an artificially short-ranged vortex interaction 
in order to arrive at a pronounced magnetization feature such as that in Fig.1. Furthermore, 
the fact that the logarithmic field derivative of the magnetization in the low field limit 
is substantially smaller than that at fields larger than $B_{\phi}$ indicates that there must be a 
second free energy contribution (i.e. entropy) that favours vortex creation and column occupation.

\subsection*{\normalsize \it 3.2 Entropy contribution to the magnetization}

The possible entropy contribution Mth to the magnetization of layered superconductors in the 
London regime, arising from vortex positional fluctuations, was considered by Bulaevskii et al. 
[23]. They derived that for large fields $B \gg B_{cr}$, $M_{th} = 
(k_{B}T/\Phi_{0}s)\ln(B_{0}/B)$, with 
$B_{0} \approx B_{c2}(\varepsilon_{0}s/k_{B}T) $
a characteristic field related to the elemental phase area. As a consequence, the logarithmic field
derivative $\partial M_{rev}^{0}/\partial \ln B$ is cancelled by 
$\partial M_{th}/\partial \ln B$ at the "crossing 
temperature" $T^{*} = T_{c}/(1 + 2k_{B}T/\varepsilon_{0}s)$;
at $T^{*}$, the reversible magnetization is field-independent. The theory was extended [24] to describe 
heavy-ion irradiated layered superconductors, which exhibit two distinct 
crossing temperatures $T_{1}^{*} $
and $T_{2}^{*}$ [16,21], in the field regimes $B << B_{\phi}$ and $B >> B_{\phi}$ respectively (see Fig.2).

\begin{figure}
		\centerline{\epsfxsize 7.5cm \epsfbox{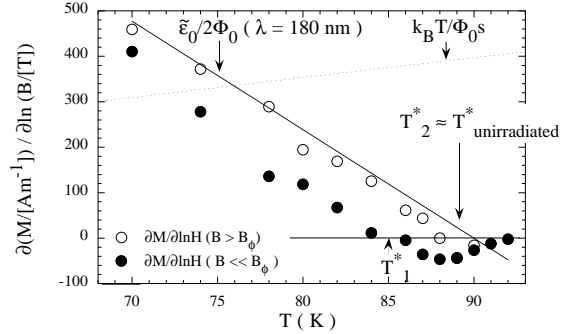}}
		\caption{\small Temperature dependence of the logarithmic field derivative of the 
		reversible magnetization for the same crystal as in Figure 1, in the field 
		regimes $B \ll B_{\phi}$ and $B \gg B_{\phi}$ respectively. The drawn lines represent the 
		value of $\partial M_{rev}^{0}/\partial \ln B = \varepsilon_{0}(0)/2\Phi_{0}$ as 
		extracted from experiment, and the 
		hypothetical entropy contribution $\partial M_{th}/\partial T 
		\partial \ln B = k_{B}T/\Phi_{0}$ [23,24].}
	\end{figure}
	
The magnitude of the entropy contribution to the magnetization can be unambiguously evaluated from 
the temperature derivative of $\partial M_{rev}/\partial \ln B$. Namely, this is expected to consist of the two contributions 
$\partial^{2} M_{rev}^{0}/\partial T \partial \ln B =$ 
\linebreak $-\varepsilon_{0}(0)/2\Phi_{0}T_{c}$ and $\partial^{2}M_{th}/\partial T 
\partial \ln B = k_{B}/\Phi_{0}s$ [25]. The first term can be read from Fig.2, 
$\partial^{2}M_{rev}^{0}/\partial T \partial \ln B \approx 23$, and yields the value of the penetration depth at 
$T=0$, $\lambda (0) = 180$ nm [26]. 
The second contribution is constant and equal to 4.4 Am$^{-1}$K$^{-1}$; the ratio 4.4/23 = 0.19 represents 
the error in $\lambda (0)$ due to the presence of fluctuations. Having

\begin{figure}
		\centerline{\epsfxsize 7.5cm \epsfbox{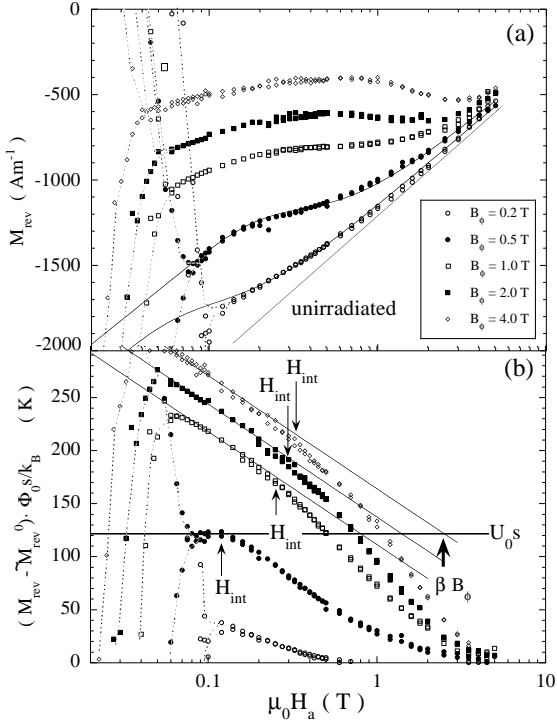}}	\vspace{-3mm}
		\caption{\footnotesize (a) Reversible magnetization of a series of lightly overdoped single crystals 
		(University of Amsterdam, $T_{c} = 83$ K), irradiated to varying matching fields, as indicated.
		The drawn lines indicate the magnetization of the unirradiated sample, and fits to the 
		data for $B_{\phi} = 0.2$ T and 0.5 T to Eq.~(1) (b) The difference between the experimental 
		magnetization and that of an unirradiated crystal, multiplied by $\Phi_{0}s$. In the low-field 
		limit, this quantity is equal to the sum of the pinning energy and the entropy per pancake. 
		The drawn lines indicate the ``bare'' pinning energy per pancake U0s (line through data for
		$B_{\phi} = 0.5$ T), and, for higher $B_{\phi} $ , the sum of $U_{0}s$ and 
		the entropy $TS = (k_{B}T/\Phi_{0}s) \ln(\beta B_{\phi} /B) $
		associated with the fact that individual pancakes belonging to the same vortex line can take 
		advantage of several columnar defect sites. The interaction field 
		$H_{int}$ is identified as that 
		at which, as a result of intervortex repulsion, pancake vortices belonging to the same stack 
		start to line up on the same columnar defect.}
	\end{figure}

	\noindent  established the value of $\lambda(0)$ with confidence, we use it to 
	evaluate the theoretical logarithmic slope 	
	$\partial M_{rev}^{0}/\partial \ln B = \varepsilon_{0}(T)/2\Phi_{0}$ (drawn in 	Fig.~2). 
	The close coincidence with the data points for $B \gg B_{\phi}$ indicates that in this field 
	regime the entropy contribution is smaller than the experimental error bar. The 
	near-equality of $M_{rev}(B \gg B_{\phi})$ with the magnetization of unirradiated crystals 
	indicates that in either case the entropy in the London regime is very modest 
	indeed, and that vortex positional fluctuations account for only a small 
	correction to the total free energy.

\begin{figure}
		\centerline{\epsfxsize 7.5cm \epsfbox{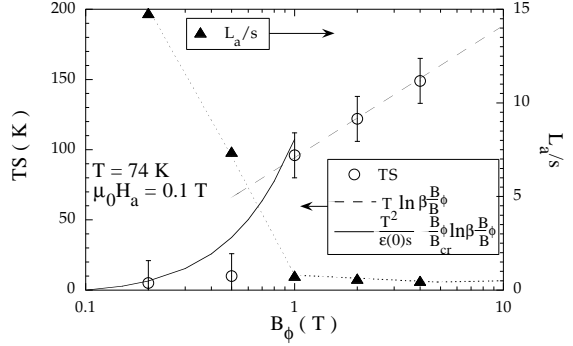}}
		\caption{\footnotesize Matching field dependence of the low-field entropy contribution to the
		free energy (measured at an applied field $\mu_{0}H_{a} = 0.1$ T). The drawn lines indicate 
		the model dependence in the regime where pancakes belonging to the same stack are
		ordered on the same column over a length $L_{a}$ ( $S \sim B_{\phi} \ln(\beta
		B_{\phi}/B)$) and that where
		vortex lines are dissociated into single pancakes occupying different columnar 
		defects ($S \sim \ln(\beta B_{\phi} /B))$. The points for $B_{\phi} = 
		0.2$ and 0.5 T are in fact upper 
		limits for $TS$, obtained from the error bar in the magnetization measurements.}
		
	\end{figure}		
		
	The situation is very different for the regime $B \ll B_{\phi}$, in which all vortices are localized 
	on a columnar defect, and where the experimental $\partial 
	M_{rev}/\partial \ln B$ lies clearly 
	below $\varepsilon_{0}(T)/2\Phi_{0}$. 
	Although the difference is smaller than the value $k_{B}T/\Phi_{0}s$ expected from theory [24], the 
	fact that it exists indicates that entropy is important in the low field regime. This is 
	further borne out by Fig.~3, which shows the reversible magnetization of a series of lightly 
	overdoped crystals with different numbers of sites available to 
	vortices, {\em i.e.} with different 
	defect density $n_{d}$ or matching field $B_{\phi}$. It appears that 
	$M_{rev}$ strongly depends on the density 
	of pinning sites at low fields ($B \ll B_{\phi}$), but not at high fields. 
	A strong $B_{\phi}$ -dependence of 
	the low-field magnetization cannot be explained in terms of pinning only, but is consistent 
	with a substantial entropy associated with the possibility for pancake vortices to occupy 
	different columnar defect sites.

	The matching-field dependence of the entropy may be extracted as follows. 
	First, subtract the magnetization of the unirradiated crystal from the experimental curves. 
	In the limit $B \ll B_{\phi}$ this yields a quantity, which, when multiplied by $\Phi_{0}s$, is exactly equal
	to the sum of the pinning energy and the entropy per pancake vortex. Next, we observe that 
	for relatively small matching fields, $B_{\phi} \leq 0.5$ T, 
	$\partial M_{rev}/\partial \ln B$ in the low-field and high-field
	limits are nearly equal. This implies that for these matching fields, the entropy is small 
	with respect to the pinning energy over the entire investigated field range. We can then confidently 
	determine the pinning energy per pancake $U_{0}s$, either by a fit to 
	a suitable expression, {\em e.g.} 
	that given in Ref.~[20], 
	
\begin{equation}
M_{rev} = M_{rev}^{0} +  \frac{U_{0}}{\Phi_{0}}  \left[ 1 - \left( 1 + 
\frac{B_{\phi}}{B} \right) \exp \left( - \frac{B_{\phi}}{B} \right) \right]      
\end{equation}

\noindent or by simple identification of the low-field plateau in the 
$B_{\phi} = 0.5$ T data of Fig.~3(b) with $U_{0}s$.
The entropy contribution to the free energy follows as the difference between the data and the pinning energy,
i.e. $TS = \Phi_{0}s(M_{rev} - M_{rev}^{0}) - U_{0}s$. It is depicted in Fig. 4. We find that the entropy is small at 
modest $B_{\phi}$, 
then rapidly rises between $B_{\phi} = 0.5$ T and 1 T, after which there 
is a further  increase proportional to $\ln B_{\phi}$.
Fig.~3 shows that in the latter regime $TS$ is proportional to $\ln B$. 

We interpret this behaviour as follows. At very small matching 
fields $B_{\phi} \ll B_{cr}$, the (attractive) 
electromagnetic and Josephson interaction between pancakes in adjacent layers prohibits vortex wandering 
between columnar defect sites; in this regime entropy can only be gained at the expense of pinning energy by 
vortex wandering into the intercolumn space [27,28]. Evidence for this entropic reduction of the pinning energy 
for very small matching fields was obtained in Refs. [7,8]. As the density 
of sites is increased (to above $6B_{cr}$,
see below), vortex lines can gain free energy by spreading to neighbouring columns. The entropy gain in this
regime occurs at the expense of elastic (shear and tilt) energy only, and can be estimated as 
$TS = k_{B}T \ln W = k_{B}T \ln N_{c} L_{s}/L_{a}$, where $L_{s}$ is the thickness of the sample in the field direction,
La is the distance over which pancakes belonging to the same vortex line remain ordered on the same columnar 
track, and $Nc \equiv  \beta B_{\phi} /B$ is the number of columns available to any one vortex. One then has 

\begin{equation}
	M_{th} = \frac{k_{B}T}{\Phi_{0}s} \frac{s}{L_{a}} \ln \left( \frac{\beta B_{\phi}}{B} 
	\right)             
	\label{eq:2}
\end{equation}

\noindent From the extrapolation of the low field data in Fig.3~(b) 
to $U_{0}s$ we obtain $\beta \sim  0.5$. 
Balancing the gain in thermal energy and the loss in tilt deformation energy 
gives $L_{a} = \varepsilon_{0}/(\gamma n_{d} k_{B}T)$ and 

\begin{equation}
	M_{th} = \frac{k_{B}T}{\Phi_{0}s} \frac{k_{B}T}{\varepsilon_{0}s} 
	\frac{B_{\phi}}{B_{cr}} \ln \left( \frac{\beta B_{\phi} }{B} \right)          
	\label{eq:3}
\end{equation}

\noindent in agreement with the rapid rise between $B_{\phi} =0.5$ T and 
$B_{\phi} = 1$ T (see Fig. 4). The typical stacklength 
can be obtained as the ratio between the experimental entropy values and 
$(k_{B}T/\Phi_{0}s)$.
As the defect density increases, $L_{a} \sim s(\varepsilon_{0} s/k_{B}T)(B_{cr}/B_{\phi})$ decreases, 
until at $B_{\phi} \gg B_{cr}(\varepsilon_{0}s/k_{B}T)$ 
it becomes equal to the layer spacing s. Experimentally this crossover occurs at a temperature dependent matching field. 
For $T = 74$ K, it is at $B_{\phi} \sim 1$ T. When the alignment length has decreased to a single pancake vortex, 
further entropy gain can only be obtained through the supplementary defect sites made available to the 
pancakes at still higher $B_{\phi}$; then, following Ref. [24], 

\begin{equation}
M_{th} = \frac{k_{B}T}{\Phi_{0}s} \ln \left(\frac{\beta B_{\phi} }{B} \right)  	
	\label{eq:4}
\end{equation}
                              
in very good agreement with the experimental data shown in Figs.~3 and 4.

The field dependence of the reversible magnetization is explained as follows. 
For the lower matching fields $B_{\phi}  \leq 0.5$ T the entropy contribution is negligible with 
respect to the pinning energy, hence the increase of $|M_{rev}|$ at 
$H_{int}$ (labelled in Fig.~3)
must be ascribed to the reduction of the total pinning energy due to intraplane pancake 
repulsion. This can be accompanied by the appearance of the first interstitial vortices. 
At higher matching fields, the low-field magnetization is the sum of 
$M_{rev}^{0}$, the gain in 
pinning energy per pancake $U_{0}s$, and the entropy gain (4) arising from the fact that each 
pancake can occupy different columns. As the field is increased, less sites are available 
per pancake, and the entropy contribution decreases logarithmically, until, 
at $H_{int}$, 
intraplane repulsion forces pancakes belonging to the same stack to lign up on the same column. 
This causes the rapid drop of $M_{th}$, which is now described by Eq. (2), and the increase of $|M_{rev}|$. 
As the matching field is approached, the increase of $|M_{rev}|$ is further enhanced by the decrease in 
pinning energy and the appearance of interstitial vortices.

\subsection*{\normalsize \it 3.4 Relation with the phase diagram}

In the following we dicuss the consequences of the pancake vortex arrangement for 
the mixed state phase diagram and vortex dynamics. Fig. 5 shows the IRL determined on 
an optimally doped Bi$_{2}$Sr$_{2}$CaCu$_{2}$O$_{8+\delta}$ crystal with 
$B_{\phi} = 0.5$ T, together with the $H_{int}$-line 
above which intraplane vortex repulsion limits the free energy gain due to columnar defect 
pinning. The $H_{int}$-line  depicted in Fig. 5 agrees very well with the ``recoupling transition''
line found in the JPR experiments of Refs. [13,14].

The irreversibility line has three well-defined parts. At low 
fields, $H_{irr} (T)$ increases exponentially 

\begin{figure}
		\centerline{\epsfxsize 7.5cm \epsfbox{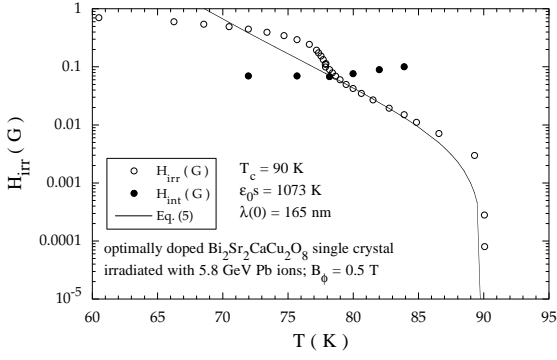}}
		\caption{\small Irreversibility line for an optimally doped 
		Bi$_{2}$Sr$_{2}$CaCu$_{2}$O$_{8}$ single crystal 
		(University of Amsterdam) irradiated with matching field $B_{\phi} = 0.5$ T. The IRL is
		determined as the temperature at which a third harmonic in the AC shielding 
		current can be first observed upon cooling. The drawn line represents a fit to 
		Eq.~(5) with the indicated parameter values.}
			\end{figure}
			
\noindent as temperature is lowered. When the interaction (or ``recoupling'') 
field $H_{int}$ is reached, 
the exponential increase stops and $H_{irr}$ rises sharply. Thus, at the precise field where 
intraplane vortex repulsion starts to determine the vortex arrangement in the columnar 
defect potential, vortex localization is spectacularly improved and vortex dynamics 
slowed down. Above Hint, every vortex line is localized not only by the defect potential 
but also by the potential created by the neighbouring vortices. As field is increased further,
$H_{irr}$ eventually starts to decrease roughly linearly with temperature, presumably because 
the proliferation of weakly-localized interstitial vortices determines flux dynamics.

The behaviour of the IRL for different matching fields is shown in Fig.~6. In the low-field,
``exponential'' regime, $H_{irr}$ increases monotonically with ion dose 
up to $B_{\phi}\sim 30$ mT, after which 
it saturates to a dose-independent exponential dependence on temperature. Such a saturation of 
the irreversibility field is reminiscent of experiments on heavy-ion irradiated 
YBa$_{2}$Cu$_{3}$O$_{7-\delta}$,
where it signals the stability limit of the entangled vortex liquid with respect to the introduction
of correlated disorder. In the low-matching field regime below the saturation (i.e. $B_{\phi} < 30$ mT),
the exponential temperature dependence of $H_{irr}$ can be observed up to 
$H_{irr} ~ B_{\phi}$, after which it crosses 
over to a linear dependence $H_{irr} \sim (1-T/Tc)$. For $B_{\phi} > 30$ mT, 
the abovementioned break at $H_{int} \sim B_{\phi}/6$
appears. It is found for the whole investigated range 0.1 T $< B_{\phi} < 
4$ T, always near the same fraction $B_{\phi}/6$.
At fields above $B_{\phi}/6$, $H_{irr}$ decreases with increasing matching field. This is

\begin{figure}
		\centerline{\epsfxsize 7.5cm \epsfbox{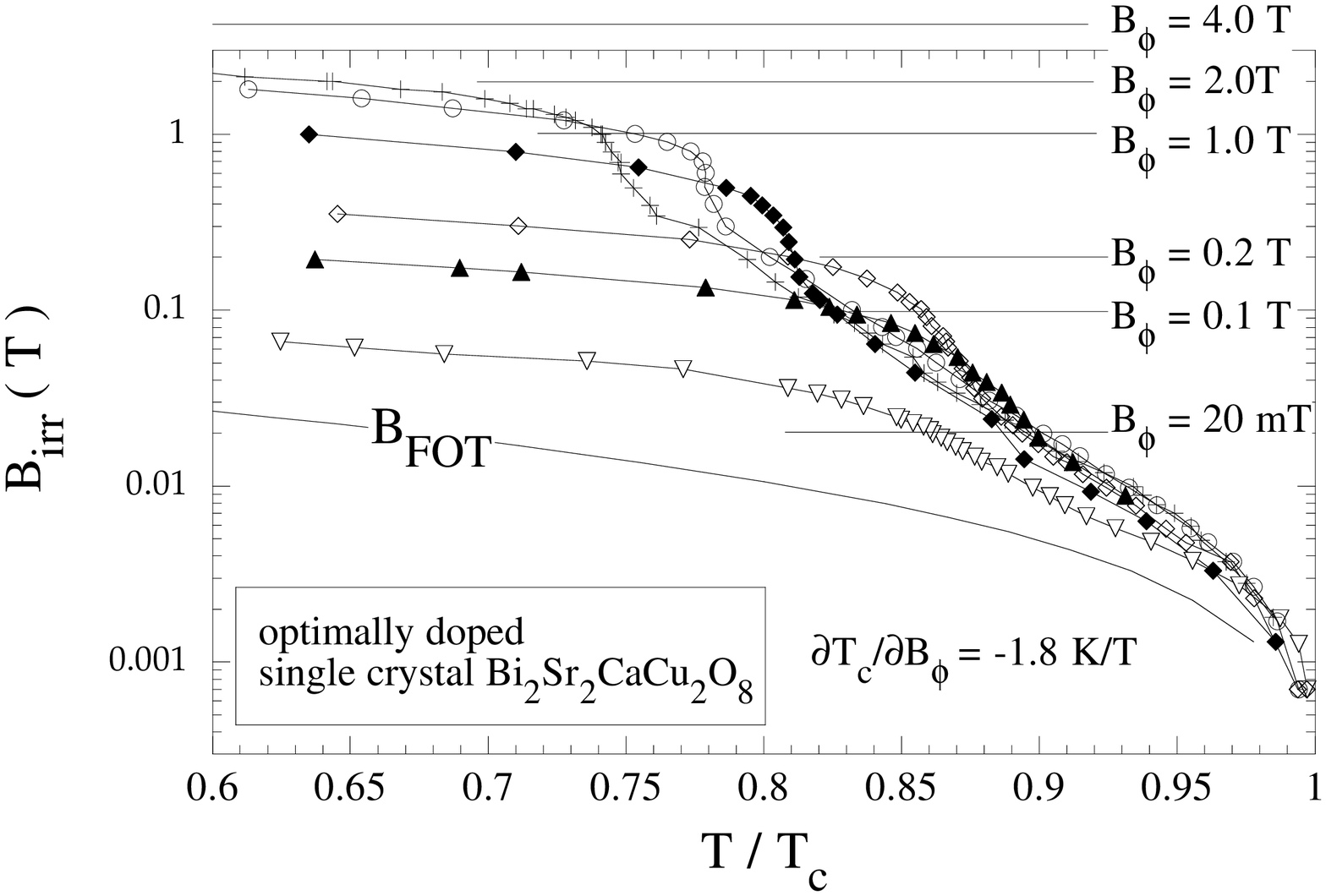}}
		\caption{Irreversibility lines for optimally doped Bi2Sr2CaCu2O8 single 
		crystals (University of Tokyo) with different matching fields $0 < 
		B_{\phi} < $ T. 
		The line labelled $B_{FOT}$ indicates the first order transition field in the pristine
		crystal. Other lines are guides to the eye.}
			\end{figure}

\noindent again the result of the fact that an increase of the number of available columnar defect 
sites promotes vortex wandering and eventually, diffusion.

\subsection*{\normalsize \it 3.5 Vortex delocalization in the low-field limit}

The low-field regime $H < H_{int}$ in which the irreversibility line saturates 
to a $B_{\phi}$-independent exponential 
temperature dependence is that in which interactions between different vortex lines are irrelevant, and, 
therefore, where every pancake vortex can find a suitable columnar defect 
site. Moreover, since for $B_{\phi} \leq 0.5$ T, 
pancakes belonging to the same vortex line are well-aligned even in the 
vortex liquid ($L_{a} \sim 10s$ at 74 K), the 
coincidence of the exponential part of the irreversibility lines for 
0.1 T$ < B_{\phi} < 4$ T implies that near $H_{irr}(T)$ 
pancakes belonging to the same stack are aligned on the same ion track irrespective of the matching field.
Vortex motion and delocalization can then be modelled in terms of a ``discrete superconductor'' in which the 
only sites allowed to the vortices are defect sites. The absence of strong positional fluctuations inferred 
in Sections 3.1 and 3.2 implies that the main thermal fluctuations will be occasional single-pancake jumps 
to neighbouring columns. These can be seen as bound vacancy-interstitial pairs (or ``quartets'' [30]), with 
separation $R$ and energy $\varepsilon_{q} = \varepsilon_{0}s(R/\gamma 
s)^{2}$, present in the vortex system of the discrete superconductor. The 
onset of vortex diffusion at the IRL corresponds to the defect pair unbinding; 
hence, $k_{B}T_{irr} \sim \varepsilon_{q}(R_{l})$ where 
$R_{l} \sim (\Phi_{0}/B)^{1/2}\exp(-\varepsilon_{0}s/2k_{B}T)$ is the typical distance between 
free dislocations (free pancakes) in the vortex 
liquid. Gathering terms, one has $k_{B}T_{irr} = \varepsilon_{0}s 
(\Phi_{0}/\gamma^{2}s^{2}) \exp(\varepsilon_{0}s/k_{B}T)$ or

\begin{equation}
B_{irr} = B_{cr} 
\frac{\varepsilon_{0}s}{k_{B}T}\exp\left(\frac{\varepsilon_{0}s}{k_{B}T}\right)   (B_{cr} \ll B < H_{int})	
	\label{eq:5}
\end{equation}

\noindent which satisfactorily describes the exponential part of the IRL, 
with parameters $\lambda (0) = 165$ nm
and $B_{cr} = 6.5 $mT (i.e. $\gamma \approx 350$) (see Fig.~5). The upper 
limit of applicability of Eq.~(5), as this emerges 
from Section 3.3 ({\em i.e.} $B < H_{int} = B_{\phi}/6$) also gives the lowest 
$B_{\phi}$ -value for which Eq.~(5) applies:
$B_{\phi} = 6 B_{cr} \sim 40$ mT, in very good agreement with the matching field value at which the IRL saturates.
We conclude that the value $B_{\phi} = 6B_{cr}$ delimits the low defect-density regime, in which thermal vortex 
excursions into the intercolumn space are allowed, from the high-defect density regime where the
``discrete superconductor''-model of strongly localized vortices is applicable.

\section*{\normalsize \bf 4. Conclusions}
Reversible magnetization measurements on Bi$_{2}$Sr$_{2}$CaCu$_{2}$O$_{8+\delta}$ single crystals containing
amorphous columnar
defects introduced by heavy-ion irradiation are used to determine the magnitude of the pinning energy 
and entropy contributions to the mixed state free energy. Whereas the pinning energy contribution can 
always be identified, the importance of entropy depends strongly on the defect  density and the magnetic 
field. For low defect  densities and unirradiated crystals, the entropy contribution to the free energy 
in the London regime ({\em i.e.} the entropy gain associated with vortex positional fluctuations) is found to 
be very minor with respect to the vortex core, electromagnetic, and pinning energy contributions. It 
therefore does not determine the main features of the phase diagram. At higher matching fields (typically 
above 1 T) and low magnetic fields $B < H_{int} ~ B_{\phi}/6$, the entropy associated with the possibility of pancakes
belonging to the same vortex line to occupy different defect sites becomes important. This configurational
entropy greatly exceeds the fluctuation entropy in unirradiated samples; it can attain values that are more 
than twice the pinning energy, for $B_{\phi} \sim 4$ T. When the magnetic 
field is increased above $H_{int}$, intraplane
pancake vortex repulsion causes the total pinning energy and the configurational entropy to diminish. 
Nevertheless, vortices become more strongly localized, since they are now retained not only by the columnar
defect  potential, but also by that of the neighbouring vortices. This effect is apparent as a steep increase
in the IRL, and as the recently found ``recoupling'' transition in JPR experiments [13,14].


\begin{thebibliography}{99}
\bibitem{Gi} Note that the ratio between these two terms is just 2Gi1/2, where Gi is the two-dimensional Ginzburg number. In optimally doped Bi2Sr2CaCu2O8+d, Gi = 0.01.
\bibitem{l} This $\lambda(0)$ value is approximately 4\% larger than that in unirradiated crystals.
Namely, the large density of columnar defects $n_{d} \gg \lambda^{-2}$ entails a reduction of the superfluid 
density and a renormalization of the vortex currents, which can be expressed by the substitution
$\lambda^{-2} \rightarrow \lambda^{-2}(1 - 2\pi c_{0}^{2}n_{d})$
\end{thebibliography}
\end{document}